# Privacy-by-Design Framework for Assessing Internet of Things Applications and Platforms

Charith Perera[1], Ciaran McCormick[1], Arosha K. Bandara[1], Blaine A. Price[1]
Bashar Nuseibeh[1,2]
[1] The Open University, Milton Keynes, UK
[2] Lero - The Irish Software Research Centre, Limerick, Ireland
firstname.lastname@open.ac.uk

## ABSTRACT
The Internet of Things (IoT) systems are designed and developed either as standalone applications from the ground-up or with the help of IoT middleware platforms. They are designed to support different kinds of scenarios, such as smart homes and smart cities. Thus far, privacy concerns have not been explicitly considered by IoT applications and middleware platforms. This is partly due to the lack of systematic methods for designing privacy that can guide the software development process in IoT. In this paper, we propose a set of guidelines, a privacy-by-design framework, that can be used to assess privacy capabilities and gaps of existing IoT applications as well as middleware platforms. We have evaluated two open source IoT middleware platforms, namely *OpenIoT* and *Eclipse SmartHome*, to demonstrate how our framework can be used in this way.

## ACM Classification Keywords
K.4.1 Computers and Society: Public Policy Issues—*Privacy*; D.2.10 Software Engineering: Design—*Methodologies*

## INTRODUCTION
The Internet of Things (IoT) [34] is a network of physical objects or *'things'* enabled with computing, networking, or sensing capabilities which allow these objects to collect and exchange data. To make IoT application development easier, a variety of IoT middleware platforms have been proposed and developed. TThese platforms offer distributed system services that have standard programming interfaces and protocols, which help solve problems associated with heterogeneity, distribution and scale in IoT applications development. These services are called 'middleware' as they sit *'in the middle'*, in a layer above the operating system and networking software and below domain-specific applications [1]. Our proposed privacy-by-design (PbD) framework can be used to assess both IoT applications and middleware platforms without any changes and agnostic to their differences. Therefore, in this paper, we use the terms 'application' and 'middleware platform' interchangeably.

Our research is motivated by a lack of privacy protection features in both IoT applications and middleware platforms. We also recognise that existing privacy-by-design frameworks do not provide specific guidance that can be used by software engineers to design IoT applications and middleware platforms. Further, recent security and privacy breaches in IoT solutions domain (e.g., Internet connected baby monitor [40]) have also motived our research.

In recent years, many parties have built IoT middleware platforms, from large corporations (e.g., Microsoft Azure IoT) to start-ups (e.g., wso2.com, Xively), from proprietary to open source (e.g., KAAproject.org), and from academic organisations (e.g., OpenIoT.eu) to broader communities (e.g., Eclipse Smart Home: eclipse.org/smarthome). Thus far, privacy has not been considered explicitly by any of these platforms, we believe partly due to a lack of privacy-by-design methods for the IoT. To address this, we propose a privacy-by-design (PbD) framework that can guide software engineers to systematically assess the privacy capabilities of IoT applications and middleware platforms. We suggest that the proposed framework can also be used to design new IoT platforms. However, in this paper, we only focus on assessing existing IoT platforms.

There are number of existing frameworks that have been proposed to help elicit privacy requirements and to design privacy capabilities in systems. The original privacy-by-design framework was proposed by Ann Cavoukian [3]. This framework identifies seven foundational principles that should be followed when developing privacy sensitive applications. These are: (1) proactive not reactive; preventative not remedial, (2) privacy as the default setting, (3) privacy embedded into design, (4) full functionality positive-sum, not zero-sum, (5) end-to-end security; full life-cycle protection, (6) visibility and

transparency- keep it open, and (7) respect for user privacy, keep it user-centric. These high level principles are proposed for computer systems in general but does not provide enough details to be adopted by software engineers when designing and developing IoT applications.

Building on the ideas of engineering privacy-by-architecture vs. privacy-by-policy presented by Spiekerman and Cranor [39], Hoepman [20] proposes an approach that identifies eight specific privacy design strategies: minimise, hide, separate, aggregate, inform, control, enforce, and demonstrate. This is in contrast to other approaches that we considered. In a similar vein, LINDDUN [11] is a privacy threat analysis framework that uses data flow diagrams (DFD) to identify privacy threats. It consists of six specific methodological steps: define the DFD, map privacy threats to DFD elements, Identify threat scenarios, prioritise threats, elicit mitigation strategies, and select corresponding privacy enhancing technologies. However, both LINDDUN and Hoepman's framework are not aimed at the IoT domain. Further, they not prescriptive enough in guiding software engineers.

In contrast, the STRIDE [21] framework was developed to help software engineers consider security threats, it is an example framework that has been successfully used to build secure software systems by industry. It suggests six different threat categories: spoofing of user identity, tampering, repudiation, information disclosure (privacy breach or data leak), denial of service, and elevation of privilege. However, its focus is mostly on security rather than privacy concerns.

On the other hand, designing IoT applications is much more difficult than designing desktop, mobile, or web applications [32]. This is beacause a typical IoT application requires both software and hardware (e.g., sensors) to work together on multiple heterogeneous nodes with different capabilities under different conditions [30]. Assessing an IoT application in order to find privacy gaps is a complex task that requires systematic guidance. For these reasons, we believe that IoT development would benefit from having a privacy-by-design framework that can systematically guide software engineers to assess (and potentially design new) IoT applications and middleware platforms. Typically, systematic guidelines will generate a consistent result irrespective of who carried out a given assessment. Such a framework will also reduce the time taken to assess a given application or platform.

**INTERNET OF THINGS: DATA FLOW**
In this section, we briefly discuss how data flows in a typical IoT application that follows a centralised architecture pattern [38]. This helps us to introduce privacy guidelines and their applicability to different types of computational nodes and data life-cycle phases. As illustrated in Figure 1, in IoT applications, data moves from sensing devices to gateway devices to a cloud infrastructure [30]. This is the most common architecture, also called centralised architecture, used in IoT application development [38]. Each of these devices have different computational capabilities. They also have different types of access to energy sources from permanent to solar power to battery power. Further, depending on the availability of knowledge, each device may have limitations as to the type of data processing that can be done. An IoT application should integrate all these different types of devices with different capabilities. We believe that IoT middleware platforms should provide in-built privacy protection capabilities. As a result, IoT application developers will be able to easily use middleware platforms to achieve their own objectives while protecting user privacy.

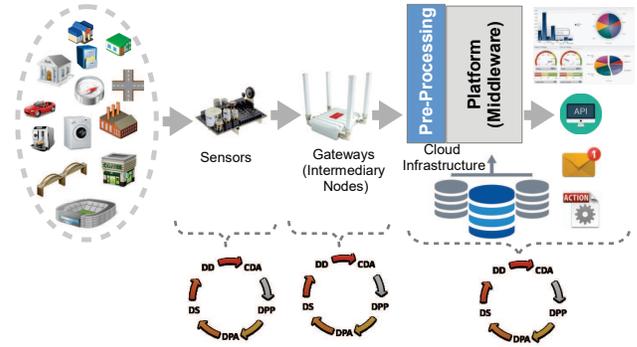

Figure 1: Typical data flow in IoT Applications

We divided the data life cycle into five phases in order to structure our discussion. Within each device (also called node), data moves through five data life cycle phases: Consent and Data Acquisition [CDA], Data Preprocessing [DPP], Data Processing and Analysis [DPA], Data Storage [DS] and Data Dissemination [DD]. The CDA phase comprises routing and data reading activities by a certain node. DPP describes any type of processing performed on raw data to prepare it for another processing procedure [36]. DPA is, broadly, *"the collection and manipulation of items of data to produce meaningful information"* [15]. DS is the storage of raw data of processed information for later retrieval and DD is the transmission of data to an external party.

We assume that all the data life cycle phases are present in all nodes in an IoT application to be utilised by software engineers to protect user privacy. However, based on the decisions taken by the engineers, some data life cycle phases in some nodes may not be utilised. For example, a sensor node may utilise the DPP phase to average temperature data. Then, without using both DPA and DS phases to analyse or store data (due to hardware and energy constraints) the sensor node may push the averaged data to the gateway node in the DD phase.

**PRIVACY-BY-DESIGN GUIDELINES**
After reviewing existing privacy design frameworks, we determined that Hoepman's [20] is the most appropriate starting point for developing a more detailed privacy-by-design framework for IoT. Additionally, Hoepman's design strategies helps us to organise and structure our paper [31]. In this section, we derive privacy-by-design guidelines by examining Hoepman's eight design strategies. These guidelines are not fool-proof recommendations that can be used without careful thought and consideration of the specific objectives, implementations, execution, and other factors specific to each IoT application or middleware platform.

It is important to note that our proposed guidelines should not be used to compare different IoT application or platforms. The primary reason is that each IoT application or platforms is designed to serve a specific purpose or category of application. For example, the *SmartHome* (eclipse.org) platform is designed to act as a home hub to enable sensing and actuation within a household environment. In contrast, *OpenIoT* [22] is designed to act as a cloud middleware for smart city applications. Therefore, they are not comparable platforms. However, if the platforms in question are very similar in terms of intended functionality, our guidelines can be used to compare them from a privacy perspective with the intention of selecting one over the other.

We developed these guidelines to act as a framework to support software engineers, so they can adopt our guidelines into their IoT applications in a customised manner. For example, certain applications will require aggregation of data from different sources to discover new knowledge (i.e. new pieces of information). We do not discourage such approaches as long as data is acquired through proper consent processes. However, IoT applications, at all times, should take all possible measures to achieve their goals with the minimum amount of data. This means that out of eight privacy design strategies proposed by Hoepman's [20], minimisation is the most important strategy. These guidelines are derived through analysis of literature and use-cases.

The relationship between Hoepman's [20] design strategies and our guidelines are presented in Table 1. Broadly, we have identified two major privacy risks, namely, secondary usage ($\otimes$) and unauthorised access ($\ominus$) that would arise as consequences of not following guidelines. The usage of already collected data for purposes that were not initially consented by the data owners can be identified as secondary usage [26]. Secondary usage can lead to privacy violations. Unauthorised access is when someone gains access to data without proper authorisation during any phase of the data life cycle. We will use the symbols shown above to denote which threat is relevant to each guideline.

**Minimise data acquisition [($\otimes$) ($\ominus$)]**
This guideline suggests to minimise the amount of data collected or requested by an IoT application [13]. Minimisation includes minimising data types (e.g., energy consumption, water consumption, temperature), minimum duration (e.g., hours, days, weeks, months), and minimum frequency (i.e., sampling rate) (e.g., one second, 30 seconds, minutes).

**Minimise number of data sources [($\otimes$) ($\ominus$)]**
This guideline suggests to minimise the number of data sources used by an IoT application. Depending on the application, it may be required to collect data from different sources. Each data source may hold information about an individual (e.g., Databox [4]). Alternatively, multiple data sources may hold pieces of information about a person [6] (e.g., Fitbit activity tracking service may hold an person's activity data while a hospital may hold his health records). Aggregation of data from multiple sources allow third parties to identify personal details that could lead to privacy violations (e.g., aggregating medical records and activity data).

**Minimise raw data intake [($\otimes$) ($\ominus$)]**
Wherever possible, IoT applications should reduce the amount of raw[1] data acquired by the system [33]. Raw data could lead to secondary usage and privacy violation. Therefore, IoT platforms should consider converting raw data into secondary context data [34]. For example, IoT applications can generate orientation (e.g., sitting, standing, walking) by processing accelerometer data, storing only the results (i.e. secondary context) and discarding the raw accelerometer values.

**Minimise knowledge discovery [($\otimes$)]**
This guideline suggests to minimise the amount of knowledge discovered within an IoT application [2]. IoT applications should only discover the knowledge necessary to achieve their primary objectives. For example, if the objective is to recommend food plans, it should not attempt to infer users' health status without their explicit permission.

**Minimise data storage [($\otimes$) ($\ominus$)]**
This guideline suggests to minimise the amount of data (i.e. primary or secondary) stored by an IoT application [39]. Any piece of data that is not required to perform a certain task should be deleted. For example, raw data can be deleted once secondary contexts are derived. Further, personally identifiable data need not be stored.

**Minimise data retention period [($\otimes$) ($\ominus$)]**
This guideline suggests to minimise the duration for which data is stored (i.e. avoid retaining data for longer than it is needed) [23]. Long retention periods provide more time for malicious parties to attempt unauthorised access to the data. Privacy risks are also increased because long retention periods could lead to unconsented secondary usage.

---
[1]Unprocessed and un-fused data can be identified as Raw data (also called *primary context*[34]). For example, X-axis value of an accelerometer can be identified as raw data. Knowledge (e.g. current activity = 'walking') generated by processing and fusing X-, Y-, and Z-axis values together can be identified as processed data (also called *secondary context*[34]).

**Hidden data routing [(⊖)]**
In the IoT, data is generated within sensor nodes. The data analysis typically happens within cloud servers. Therefore, data is expected to travel between different types of computational nodes before arriving at the processing cloud servers. This type of routing reveals a user's location and usage from anyone conducting network surveillance or traffic analysis. To makes it more difficult for Internet activities to be traced back to the user, this guideline suggests that IoT applications should support and employ anonymous routing mechanisms (e.g., Tor [24]).

**Data anonymisation [(⊗) (⊖)]**
This guideline suggests to remove personally identifiable information before the data gets used by the IoT application so that the people described by the data remain anonymous. Removal of personally identifiable information reduces the risk of unintended disclosure and privacy violations [12].

**Encrypted data communication [(⊖)]**
This guideline suggests that different components in an IoT application should consider encrypted data communication wherever possible [16]. Encrypted data communication would reduce the potential privacy risks due to unauthorised access during data transfer between components. There are multiple data communication approaches based on the components involved in an IoT application, namely, 1) device-to-device, 2) device-to-gateway, 3) device-to-cloud, and 4) gateway-to-cloud. Sensor data communication can be encrypted using symmetric encryption AES 256 [8] in the application layer. Typically, device-to-device communications are encrypted at the link layer using special electronic hardware included in the radio modules [16]. Gateway-to-cloud communication is typically secured through HTTPS using Secure Sockets Layer (SSL) or Transport Layer Security (TLS).

**Encrypted data processing [(⊖)]**
This guideline suggests to process data while encrypted. Encryption is the process of encoding data in such a way that only authorised parties can read it. However, sometimes, the party who is responsible for processing data should not be allowed to read data. In such circumstances, it is important to process data in encrypted form. For example, homomorphic encryption [14] is a form of encryption that allows computations to be carried out on cipher-text, thus generating an encrypted result which, when decrypted, matches the result of operations performed on the plain-text [17] .

**Encrypted data storage [(⊖)]**
This guideline suggests that IoT applications should store data in encrypted form [5]. Encrypted data storage reduces any privacy violation due to malicious attacks and unauthorised access. Data encryption can be applied at different levels from sensors [18] to the cloud.

Depending on the circumstances, data can be encrypted using both hardware and software technologies.

**Reduce data granularity [(⊗)]**
The granularity is the level of depth represented by the data. High granularity refers to an atomic grade of detail and low granularity zooms out into a summary view of data [28]. For example, dissemination of location can be considered as coarse-grained and full address can be considered as fine-grained. Therefore, releasing fine grained information always has more privacy risks than coarse-grained data as they contain more information. Data granularity has a direct impact on the quality of the data as well as the accuracy of the results produced by processing such data [41]. IoT applications should request the minimum level of granularity that is required to perform their primary tasks. Higher level of granularity could lead to secondary data usage and eventually privacy violations.

**Query answering [(⊗)]**
This guideline suggests to release high-level answers to queris when disseminating information, without releasing raw data. For example, a sample query would be *'how energy efficient a particular household is?'* where the answer would usa a 0-5 scale. Raw data can always lead to privacy violations due to secondary usage. One such implementation is openPDS/SafeAnswers [10] which allows users to collect, store, and give high level answers to the queries while protecting their privacy.

**Repeated query blocking [(⊗)]**
This guideline goes hand-in-hand with the Query answering guideline. When answering queries, IoT applications need to make sure that they block any malicious attempts to discover knowledge that violates user privacy through repeated queries (e.g., analysing intersections of multiple results) [35].

**Distributed data processing [(⊗) (⊖)]**
This guideline suggests that an IoT application should process data in a distributed manner. Similar, approaches are widely used in the wireless sensor network domain [37]. Distributed processing avoids centralised large-scale data gathering. As a result, it deters any unauthorised data access attempts. Different types of distributed IoT architectures are discussed in [38].

**Distributed data storage [(⊗) (⊖)]**
This guideline recommends storing data in a distributed manner [29]. Distributed data storage reduces any privacy violation due to malicious attacks and unauthorised access. It also reduces privacy risks due to unconsented secondary knowledge discovery.

**Knowledge discovery based aggregation [(⊗)]**
Aggregation of information over groups of attributes or groups of individuals, restricts the amount of detail in the personal data that remains [20]. This guideline suggests to discover knowledge though aggregation and replace raw data with discovered new knowledge. For example, *'majority of people who visited the park on [particular date] were young students'* is an aggregated result that is sufficient (once collected over a time period) to perform further time series based sales performance analysis of a near-by shop. Exact timings of the crowd movements are not necessary to achieve this objective.

**Geography based aggregation [(⊗)]**
This guideline recommends to aggregate data using geographical boundaries [27]. For example, a query would be *'how many electric vehicles used in each city in UK'*. The results to this query would be an aggregated number unique to the each city. It is not required to collect or store details about individual electric vehicles.

**Chain aggregation [(⊗)]**
This guideline suggests to perform aggregation on-the-go while moving data from one node to another. For example, if the query requires a count or average, this can be done without pulling all the data items to a centralised location. Data will be sent from one node to another until all the nodes get a chance to respond. Similar techniques are successfully used in wireless sensor networks [25]. This type of technique reduces the amount of data gathered by a centralised node (e.g., cloud server). Further, such aggregation also eliminates raw data from the results, thus reducing the risk of secondary data usage.

**Time-Period based aggregation [(⊗)]**
This guideline suggests to aggregate data over time (e.g., days, week, months) [9]. This reduces the granularity of data and also reduces the secondary usage that could lead to privacy violations. For example, energy consumption of a given house can be acquired and represented in aggregated form as 160 kWh per month instead of on a daily or hourly basis.

**Category based aggregation [(⊗)]**
Categorisation based aggregation approaches can be used to reduce the granularity of the raw data [9]. For example, instead of using exact value (e.g., 160 kWh per month), energy consumption of a given house can be represented as 150-200 kWh per month. Time-Period based and category based aggregation can be combined together to reduce data granularity.

**Information Disclosure [(⊗)]**
This guideline suggests that data subjects should be adequately informed whenever data they own is acquired, processed, and disseminated. The 'Inform' step can take place at any stage of the data life cycle and can be broadly divided into two categories: pre-inform and post-inform. Pre-inform takes place before data enters to a given data life cycle phase. Post-inform takes place soon after data leaves a given data life cycle phase [19].

- *Consent and Data Acquisition:* what is the purpose of the data acquisition?, What types of data are requested?, What is the level of granularity?, What are the rights of the data subjects?

- *Data Pre-Processing:* what data will be taken into the platform?, what data will be thrown out?, what kind of pre-processing technique will be employed?, what are the purposes of pre-processing data?, what techniques will be used to protect user privacy?

- *Data Processing and Analysis:* what type of data will be analysed?, what knowledge will be discovered?, what techniques will be used?.

- *Data Storage:* what data items will be stored? how long they will be stored? what technologies are used to store data (e.g. encryption techniques)? is it centralised or distributed storage? will there be any back up processes?

- *Data Dissemination:* with whom the data will be shared? what rights will receivers have? what rights will data subjects have?

**Control [(⊗)]**
This guideline recommends providing privacy control mechanisms for data subjects [9]. Control mechanisms will allow data owners to manage data based on their preference. There are different aspects that the data owner may like to control. However, controlling is a time consuming task and not every data owner will have the expertise to make such decisions.

Therefore, it is a software engineer's responsibility to determine the kind of controls that are useful to data owners in a given IoT application context. Further, it is important to provide some kind of default set of options for data owners to choose from, specially in the cases where data subjects do not have sufficient knowledge. Some potential aspects of control are 1) data granularity, 2) anonymisation technique, 3) data retention period, 4) data dissemination.

**Logging [(⊗) (⊖)]**
This guideline suggests to log events during all phases [9]. It allows both internal and external parties to examine what has happened in the past to make sure a given system has performed as promised. Logging could include but is not limited to event traces, performance parameters, timestamps, sequences of operations performed over data, any human interventions. For example, a log may include the timestamps of data arrival, operations performed in order to anonymise data, aggregation techniques performed, and so on.

**Auditing**
This guideline suggests to perform systematic and independent examinations of logs, procedures, processes, hardware and software specifications, and so on [9]. The logs above could play a significant role in this process.

Table 1: Analysis of Privacy-by-Design Framework

| Influence | Guideline | DA | DPP | DPA | DS | DD | Minimize | Hide | Separate | Aggregate | Inform | Control | Enforce | Demonstrate | Privacy Threats |
|---|---|---|---|---|---|---|---|---|---|---|---|---|---|---|---|
| [13] | 1-Minimise data acquisition | ✓ | ✓ | | | | ✓ | ✓ | | | | | | | ⊗ ⊖ |
| [6] | 2-Minimise number of data sources | ✓ | | | | | ✓ | | | | | | | | ⊗ ⊖ |
| [33] | 3-Minimise raw data intake | ✓ | ✓ | | | | ✓ | | | ✓ | | | | | ⊗ ⊖ |
| [2] | 4-Minimize knowledge discovery | | | ✓ | | | ✓ | | | | | | | | ⊗ |
| [39] | 5-Minimize data storage | | | | ✓ | | ✓ | | | | | | | | ⊗ ⊖ |
| [23] | 6-Minimize data retention period | | | | ✓ | | ✓ | ✓ | | | | | | | ⊗ ⊖ |
| [24] | 7-Hidden data routing | ✓ | | | | ✓ | | ✓ | | | | | | | ⊖ |
| [12] | 8-Data anonymization | ✓ | ✓ | ✓ | | ✓ | | ✓ | | | | | | | ⊗ ⊖ |
| [16] [8] | 9-Encrypted data communication | ✓ | | | | ✓ | | ✓ | | | | | | | ⊖ |
| [14] [17] | 10-Encrypted data processing | | ✓ | ✓ | | | | ✓ | | | | | | | ⊖ |
| [5] [18] | 11-Encrypted data storage | | | | | | | ✓ | | | | | | | ⊖ |
| [41] | 12-Reduce data granularity | ✓ | ✓ | ✓ | | ✓ | | ✓ | | | | | | | ⊗ |
| [10] | 13-Query answering | | | | | ✓ | ✓ | ✓ | | | | | | | ⊗ |
| [35] | 14-Repeated query blocking | | | | | ✓ | ✓ | ✓ | | | | | | | ⊗ |
| [37] [38] | 15-Distributed data processing | | | ✓ | | | | | ✓ | | | | | | ⊗ ⊖ |
| [29] | 16-Distributed data storage | | | | ✓ | | | | ✓ | | | | | | ⊗ ⊖ |
| [20] | 17-Knowledge discovery | ✓ | ✓ | ✓ | ✓ | ✓ | | | | ✓ | | | | | ⊗ |
| [27] | 18-Geography based aggregation | ✓ | ✓ | ✓ | ✓ | ✓ | | | | ✓ | | | | | ⊗ |
| [25] | 19-Chain aggregation | ✓ | ✓ | ✓ | ✓ | ✓ | | | | ✓ | | | | | ⊗ |
| [9] | 20-Time-Period based aggregation | ✓ | ✓ | ✓ | ✓ | ✓ | | | | ✓ | | | | | ⊗ |
| [9] | 21-Category based aggregation | ✓ | ✓ | ✓ | ✓ | ✓ | | | | ✓ | | | | | ⊗ |
| [19] | 22-Information Disclosure | ✓ | ✓ | ✓ | ✓ | ✓ | | | | | ✓ | | | ✓ | ⊗ |
| [9] | 23-Control | ✓ | ✓ | ✓ | ✓ | ✓ | | | | | | ✓ | ✓ | | ⊗ |
| [9] | 24-Logging | ✓ | ✓ | ✓ | ✓ | ✓ | | | | | | | | ✓ | ⊗ ⊖ |
| [9] | 25-Auditing | | | | | | | | | | | | | ✓ | |
| | 26-Open Source | | | | | | | | | | | | | ✓ | |
| | 27-Data Flow | | | | | | | | | | | | | ✓ | |
| [42] | 28-Certification | | | | | | | | | | | | | ✓ | |
| [7] | 29-Standardization | | | | | | | | | | | | | ✓ | |
| [9] | 30-Compliance | | | | | | | | | | | | | ✓ | ⊗ ⊖ |

Risk Types: Secondary Usage (⊗), Unauthorised Access (⊖)

Non-disclosure agreements may be helpful to allow auditing some parts of the classified data analytics processes.

**Open Source**
Making source code of an IoT application open allows any external party to review code. Such reviews can be used as a form of compliance demonstration. This allows external parties to verify and determine whether a given application or platform has taken sufficient measures to protect user privacy.

**Data Flow**
Data flow diagrams (e.g., Data Flow Diagrams used by Unified Modelling Language) allow interested parties to understand how data flows within a given IoT application and how data is being treated. Therefore, DFDs can be used as a form of a compliance demonstration.

**Certification**
In this context, certification refers to the confirmation of certain characteristics of an system and process. Typically, certifications are given by a neutral authority. Certification will add trustworthiness to IoT applications. TRUSTe (truste.com) [42] Privacy Seal is one example, even though none of the existing certifications are explicitly designed to certify IoT applications.

**Standardisation**
This guideline suggests to follow standard practices as a way to demonstrate privacy protection capabilities. Industry wide standards (e.g. AllJoyn allseenalliance.org) typically inherit security measures that would reduce some privacy risks as well. This refers to the process of implementing and developing technical standards. Standardisation can help to maximise compatibility, interoperability, safety, repeatability, or quality. Standardisation will help external parties to easily understand the inner workings of a given IoT application [7].

**Compliance [(⊗) (⊖)]**
Based on the country and region, there are number of policies, laws and regulations that need to be adhered to. It is important for IoT applications to respect guidelines. Some regulatory efforts are ISO 29100 Privacy framework, OECD privacy principles, and European Commission Protection of personal data.

**EVALUATION**
In this section, we demonstrate how our proposed PbD framework can be used to evaluate existing IoT applications and platforms in order to find gaps in privacy. For evaluation, we used two open source IoT platforms that have been developed to accomplish different goals.

We intentionally selected two IoT platforms instead of IoT applications due to their open source nature and availability to the research community.

The OpenIoT [22] middleware infrastructure supports flexible configuration and deployment of algorithms for collecting, and filtering information streams stemming from internet-connected objects, while at the same time generating and processing important business/application events. It is more focused on enterprise and industrial IoT applications.

Eclipse Smart Home (eclipse.org/smarthome) is a middleware platform for integrating different home automation systems and technologies into one single solution that allows over-arching automation rules and uniform user interfaces. It allows building smart home solutions that have a strong focus on heterogeneous environments. As the name implies, it mainly focused on smart home (or smart office) based solutions.

**Methodology**

Here we provide a step-by-step description of how our proposed privacy-by-design framework can be used to assess IoT applications and middleware platforms. In this work, we only focus on assessing two IoT platforms, even though the proposed framework can also support designing brand new IoT applications and middleware platforms as well. We believe the following method helps software engineers to efficiently and effectively use our framework, although this is not the only way to do so.

- **Step 1:** First, software engineers need to identify how data flows in the existing system. The objective is to identify the physical devices through which data transits at runtime. However, only the categories of devices need to be identified. This is a decision that software engineers need to make. An example layout is show in Figure 2, where we have illustrated two different gateway devices to highlight the fact that we are only interested in categories of devices. Device category is typically based on the similarities in terms of computational capabilities (e.g., CPU, RAM, etc.).

- **Step 2:** Build a table for each node where columns represent data life cycle phases and rows represent each privacy-by-design guideline. In this paper, our aim is to assess two IoT middleware platforms that are designed to be hosted in an server node. We required only one table for each assessment[2]. However, when assessing an IoT application, there could be many nodes involved. In such situations, a table (e.g., Table 2) is required for each node.

- **Step 3:** Finally, software engineers go through each guideline and use the colour codes proposed below to assess their platforms. We conducted our own assessment using this approach and results are presented in Table 2. A software engineer may write notes to justify their decisions (i.e. choice of color) as well as to clarify the rationale on each cell. For example, encrypted data processing may not be possible to perform in certain categories of devices such as gateways due to their lack of computational resources . Such notes are useful when working as a team so everyone knows the rationale behind design choices.

Our proposed color coding is as follows: When a particular guideline is not applicable for a given life cycle phase, it is marked as NOT-APPLICABLE (■). If a particular guideline if fully supported by a given IoT application or platform, it is marked as FULLY-SUPPORTED (■). This means that the platform has already taken steps to protect user privacy as explained in the principles. If a particular guideline is not supported at all or it requires substantial effort to support a certain functionality, it is marked as NO-SUPPORT (■). This means that the platform has not taken necessary the steps to protect user privacy as explained in the guideline and it is either impossible or requires significant effort to fix this. When a certain guideline is not supported by a platform but provides a mechanism (e.g., plug-in architecture) to support user privacy protection through extensions, we identify them as EXTENDIBLE (■).

---

[2]Two independent assessment tables are merged due to space limitations

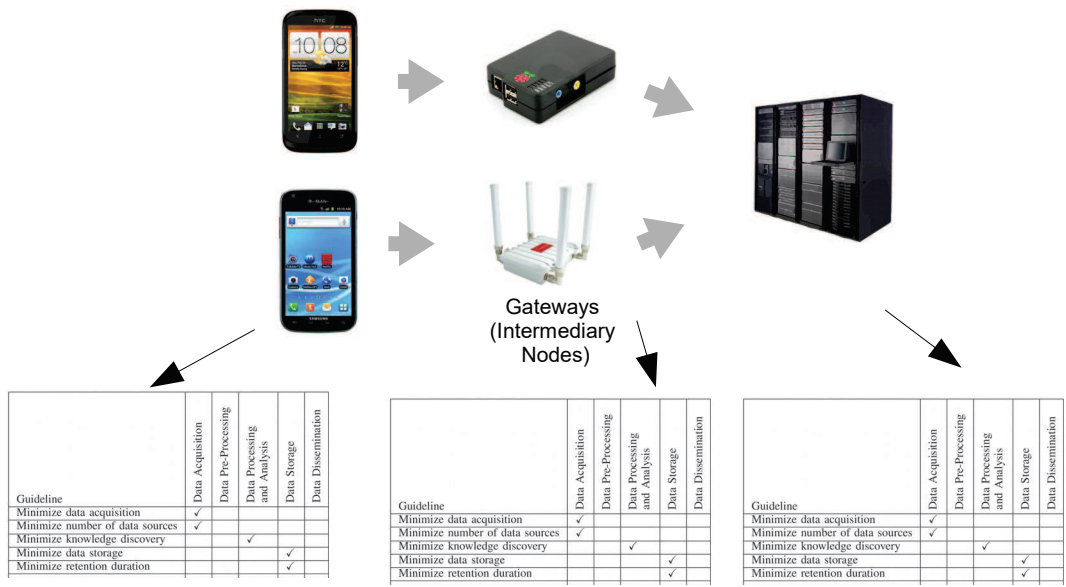

Figure 2: Evaluation Methodology

**Discussion**
Here we discuss a few of our guidelines to demonstrate how a software engineer may use our framework to evaluate their IoT applications from a privacy perspective. Our intention is not to discuss and justify how we evaluated both platforms against each guideline, but to exemplify the thought processes behind this evaluation.

Let us consider guideline #1 (Minimize Data Acquisition). This guideline can be satisfied by extending both *OpenIoT* and Eclipse *SmartHome* platform in the CDA phases. *OpenIoT* has a plug-in architecture called 'wrappers' [22]. *SmartHome* also has similar architecture called 'bundles'. These plug-ins can be easily extended to request minimum amount of data. However, such functionality is not readily available in these platforms. Therefore, we marked CDA phase as EXTENDIBLE (■).

The minimize data acquisition function is readily available in *OpenIoT* [22] in the DPP phases. It provides a mechanism to configure parameters such as 'sampling rate' using a declarative language. Therefore we marked *OpenIoT's* DPP phase as Fully Supported (■). However, no similar functionality is provided in the *SmartHome* platform.

The guidelines #25 which focused on Auditing is marked as NO-SUPPORT (■) for both platforms. The reason is that, though both platforms are open source, neither of them are audited from a privacy perspective. Due to their open source nature, code bases are regularly reviewed and audited to make sure coding standards are met. However, privacy aspects are not reviewed in current auditing sessions. We conducted similar examinations with respect to all guidelines.

**CONCLUSIONS AND FUTURE WORK**
In this paper, we presented set of guidelines, as the core of a conceptual framework, that incorporates privacy-by-design principles to guide software engineers in the systematic assessment of the privacy capabilities of Internet of Things applications and platforms. We demonstrated how our framework can be used to assess two open source IoT platforms namely, Eclipse Smart Home and OpenIoT. We also explained the step by step process of how to use our framework efficiently. The proposed summarizing technique may be helpful when software engineers need to report current statuses of their IoT applications from a privacy perspective and justify investments towards certain privacy features. Further, detailed analysis of privacy gaps will help software engineers to share their thought processes with colleagues towards design and development of new privacy features.

Table 2: Summarized Privacy Gaps Assessment for (a) *Eclipse SmartHome* and (b) *OpenIoT*

| Guideline | (a) Eclipse SmartHome Platform | | | | | (b) OpenIoT Platform | | | | |
|---|---|---|---|---|---|---|---|---|---|---|
| | CDA | DPP | DPA | DS | DD | CDA | DPP | DPA | DS | DD |
| 1-Minimise data acquisition | ■ | | | | | ■ | ■ | | | |
| 2-Minimise number of data sources | ■ | | | | | ■ | | | | |
| 3-Minimise raw data intake | ■ | ■ | | | | ■ | ■ | | | |
| 4-Minimize knowledge discovery | | ■ | ■ | | | | | ■ | | |
| 5-Minimize data storage | | | ■ | ■ | | | | | ■ | |
| 6-Minimize data retention period | | | | ■ | | | | | ■ | |
| 7-Hidden data routing | ■ | | | | ■ | ■ | | | | ■ |
| 8-Data anonymization | | ■ | | | | | ■ | | | |
| 9-Encrypted data communication | ■ | ■ | | | ■ | ■ | ■ | | | ■ |
| 10-Encrypted data processing | | ■ | ■ | | | | ■ | ■ | | |
| 11-Encrypted data storage | | | | ■ | | | | | ■ | |
| 12-Reduce data granularity | ■ | ■ | | | | ■ | ■ | | | ■ |
| 13-Query answering | | | ■ | | | | | ■ | | ■ |
| 14-Repeated query blocking | | | ■ | | | | | ■ | | ■ |
| 15-Distributed data processing | | | ■ | | | | | ■ | | |
| 16-Distributed data storage | | | | ■ | | | | | ■ | |
| 17-Knowledge discovery based aggregation | ■ | ■ | ■ | ■ | ■ | ■ | ■ | ■ | ■ | ■ |
| 18-Geography based aggregation | ■ | ■ | ■ | ■ | ■ | ■ | ■ | ■ | ■ | ■ |
| 19-Chain aggregation | ■ | ■ | ■ | ■ | ■ | ■ | ■ | ■ | ■ | ■ |
| 20-Time-Period based aggregation | ■ | ■ | ■ | ■ | ■ | ■ | ■ | ■ | ■ | ■ |
| 21-Category based aggregation | ■ | ■ | ■ | ■ | ■ | ■ | ■ | ■ | ■ | ■ |
| 22-Information Disclosure | ■ | ■ | ■ | ■ | ■ | ■ | ■ | ■ | ■ | ■ |
| 23-Control | ■ | ■ | ■ | ■ | ■ | ■ | ■ | ■ | ■ | ■ |
| 24-Logging | ■ | ■ | ■ | ■ | ■ | ■ | ■ | ■ | ■ | ■ |
| 25-Auditing | ■ | ■ | ■ | ■ | ■ | ■ | ■ | ■ | ■ | ■ |
| 26-Open Source | ■ | ■ | ■ | ■ | ■ | ■ | ■ | ■ | ■ | ■ |
| 27-Data Flow Diagrams | ■ | ■ | ■ | ■ | ■ | ■ | ■ | ■ | ■ | ■ |
| 28-Certification | ■ | ■ | ■ | ■ | ■ | ■ | ■ | ■ | ■ | ■ |
| 29-Standardization | ■ | ■ | ■ | ■ | ■ | ■ | ■ | ■ | ■ | ■ |
| 30-Compliance | ■ | ■ | ■ | ■ | ■ | ■ | ■ | ■ | ■ | ■ |

In the future, we will conduct empirical studies by recruiting software engineers to assess the privacy capabilities of open source IoT platforms with and without using our framework. Such studies will help us to derive more insights on its value in real-world settings. Further, through empirical studies, we will explore how our framework may be used by non specialised IT professionals (e.g., final year students, new software engineering graduates) to assess the existing privacy capabilities of IoT middleware frameworks.

We also plan to demonstrate how our framework can be used to design brand new IoT applications and platforms. Specifically, we will ask participants to design IoT applications to satisfy few different use case scenarios with and without our guidelines. We will measure the effectiveness of their designs using quantitative techniques. Furthermore, to help software engineers better, we are also planning to extract, design and document privacy patterns that can be easily adopted into IoT application design processes.

**ACKNOWLEDGMENTS**

We acknowledge the financial support of European Research Council Advanced Grant 291652 (ASAP).